\documentclass[journal,10pt,draftcls,onecolumn]{IEEEtran}
\pdfoutput=1

\usepackage[mathlines]{lineno}
\usepackage[hidelinks]{hyperref}
\modulolinenumbers[1]
\usepackage{amssymb}
\usepackage[]{units}
\usepackage{amsmath}
\usepackage{booktabs}
\usepackage{caption}
\usepackage{subfigure}
\usepackage{multirow}
\usepackage{float}
\usepackage{cancel}
\usepackage{listings}
\usepackage{adjustbox}
\usepackage{verbatim}

\author{
    \IEEEauthorblockN{Marco Barbieri\IEEEauthorrefmark{1}, Leonardo Brizi\IEEEauthorrefmark{1}, Enrico Giampieri\IEEEauthorrefmark{1}, Francesco Solera\IEEEauthorrefmark{2}, Gastone Castellani\IEEEauthorrefmark{1}, Claudia Testa\footnote{Dr. Claudia Testa is corresponding author, Email: claudia.testa@unibo.it}\IEEEauthorrefmark{1} and Daniel Remondini\IEEEauthorrefmark{1}}
    
    \IEEEauthorblockA{\IEEEauthorrefmark{1} Department of Physics and Astronomy, University of Bologna, Italy}\\
    \IEEEauthorblockA{\IEEEauthorrefmark{2} Deep Vision Consulting, Modena, Italy}\\
   
}

\begin{document}

\title{
Circumventing the Curse of Dimensionality in Magnetic Resonance Fingerprinting through a Deep Learning Approach
}
\maketitle

\begin{abstract}
MR fingerprinting (MRF) is a rapid growing approach for fast quantitave MRI. A typical drawback of dictionary-based MRF is its explosion in size as a function of the number of reconstructed parameters, according to the curse of dimensionality.
Deep Neural Networks (NNs) have been proposed as a feasible alternative, but these approaches are still in their infancy.

We tested different NN pipelines on simulated data: we studied optimal training procedures by including different strategies of noise addition and parameter space sampling, to achieve better accuracy and robustness to noise.
Four MRF sequences were considered, two of them designed to be more specific for $B_1^+$ parameter encoding: IR-FISP, IR-FISP-$B_1$, bSSFP and  IR-bSSFP-$B_1$. 
A comparison between NN and the dictionary approaches was performed using a numerical brain phantom.

Results demonstrated that training with random sampling and different levels of noise variance yielded the best performance. 
NN performance was greater or equal than dictionary-based approach in reconstructing MR parameter maps: the difference in performance increased with the number of estimated parameters, because the dictionary method suffers from the coarse resolution of the MR parameter space sampling.
The NN approach resulted more efficient in terms of memory and computational burden, and thus has great potential in large-scale MRF problems.

\end{abstract}

\section{Introduction}
Magnetic Resonance Fingerprinting (MRF) is a fast quantitative MRI technique able to obtain multi-parametric maps in one-shot measurement \cite{Nature2013}; many applications of the technique have been investigated since its birth, ranging from brain imaging \cite{LemassonPannetierCoqueryEtAl2016} to body MRI \cite{ChenJiangPahwaEtAl2016}, \cite{doi:10.1148/radiol.2018180836}.

The key concept of MRF is to apply a train of RF pulses with flip angle (FA) and repetition time (TR) varying according to a pattern designed to differentiate as much as possible a signal response for different tissues, so that for each voxel a so called fingerprint of the tissue is acquired.
The original methodology is based on the template matching of the experimental fingerprint with a precomputed dictionary of simulated signal evolutions, and the dot product is usually adopted as a fast similarity measure.
Dictionary exploding size, matching accuracy, robustness to noise and under sampling artifacts are the main challenges to face to bring MRF into clinical applications. The more parameters are encoded into the pulse sequence simulations ($T_1$, $T_2$, $B_0$, $B_1^+$, etc...) the bigger the size of the precomputed dictionary has to be.
However, confounding factors such as $B_1^+$ field inhomogeneities are known to be a source of artifacts in MRI \cite{JinghuaWeihuaMaolinEtAl}, and researchers are taking into account these confounding factor also in MRF to improve parameters estimation \cite{BuonincontriSawiak2015, MRF_LW_SM_2018}.
However, big dictionaries are hard to handle because they are costly both in memory usage efficiency and in computational time for the matching procedure, with size up to 150 GB \cite{MRF_LW_SM_2018}. This problem has driven the MRF community forward methods aiming to compress the size of the dictionaries by both applying Single Value Decomposition (SVD) to generate low rank approximation of the fingerprinting signals, and reducing the entries of the dictionaries by using a polynomial fitting \cite{McGivneyPierreMaEtAlDec.2014, FLOR_MRF, MingruiDanYunEtAl2017}. Nevertheless, these methods make an additional approximation, which can affect accuracy.

A strategy to overcome the limitations of the dictionary based template matching is taking advantage of Machine Learning algorithms to learn a model able to predict, after a supervised training procedure, the MR parameters given the experimental fingerprint as input. Among these, Neural Networks (NN) algorithm is a feasible tool to accomplish such a task, as NNs have been demonstrated to be universal function approximators given enough train data and model complexity \cite{Hornik1991}. 
Hence, a sufficiently large NN can learn the Inverse Transfer Function (ITF) to map the acquired MRF signal into the MR parameter space. Moreover, once the NN is trained, the prediction operation has many advantages with respect to the template matching, such as being computationally efficient, because no exhaustive search has to be performed among a dictionary. 
More importantly, it is able to predict MR parameters of unknown signals, whereas template matching approximates the MR parameters prediction to those present in the dictionary. This feature can limit quantization artefacts that can arise from dictionary approach.
Depending on the task and the architecture, NN can require thousands of training data, which in case of MRI can be both expensive and time-consuming. However, for application to MRF the NN model can be trained using simulated data, which are easy to produce by using a simulator. Simulating the data is the original way in which dictionaries are produced in MRF, and the reliability of them has been widely demonstrated by the literature \cite{Nature2013}. 
Few works already investigated the feasibility of applying Neural Networks to MRF, both with numerical simulations \cite{2017arXiv170700070V,Hoppe2017DeepLF} and both with phantoms and in-vivo acquisitions \cite{OuriBoS.2018,SpatioTemporal_CNN}.
The results of these works have demonstred the feasibility of NN approaches using both fully connected and convolutional neural networks. However, the focus of these works was primarily on demonstrating the feasibility of NN approach, and less attention has been given to which pipeline should be used to make the Neural Network model to learn an accurate and robust ITF for a given a MRF pulse sequence. For example, Virtue et al \cite{2017arXiv170700070V}  created a training data set by sampling randomly the MR parameter space, whereas Hoppe et al \cite{Hoppe2017DeepLF} and Cohen et al \cite{OuriBoS.2018} used a grid sampling. Moreover, different strategies have been applied for taking into account the noise, such as in \cite{2017arXiv170700070V,Hoppe2017DeepLF} where no noise was added in the training procedure, whereas in \cite{OuriBoS.2018} white Gaussian noise with zero mean and $1\%$ standard deviation was added during the training phase to promote robustness to noise.

The aim of this work, carried out by means of simulations, is to set and test different pipelines to apply Deep Neural Networks to MRF data to learn an accurate and robust ITF for a given MRF pulse sequence. In particular this work investigates the generalization capability of NNs in relation to the method used to sample the parameter space, either uniform random sampling or grid sampling. Then, noise robustness is increased applying different data augmentation strategies, adding white Gaussian noise to MRF training signal examples.
Once found a suitable pipeline to train the NN, a comparison between the NN approach and standard approach based on the dictionary matching in reconstructing brain quantitative maps is reported by means of a numerical simulations based on brain MR parameter values.

To show the ability of generalization of the NN approach, and how its performance scale with the number of predicted MR parameters in comparison with the dictionary approach, the NN has been applied to different MRF pulse sequences, i.e. IR balanced Steady State Precession (IR-bSSFP) \cite{Nature2013}, IR Fast Imaging with Steady state Precession (IR-FISP)\cite{YunDanNicoleEtAl2014} and its variant for $B_1^+$ estimation \cite{BuonincontriSawiak2015}. Moreover, a new MRF bSSFP sequence is here proposed to simultaneously estimate $T_1$, $T_2$, $B_0$ off-resonances and $B_1^+$ field inhomogeneities.
This procedure differs from the one presented in \cite{MRF_LW_SM_2018}, where the $B_1^+$ is measured with a standard MRI protocol independently from the MRF acquisition, and different MRF dictionaries encoding $T_1$, $T_2$ and $B_0$ are computed for each $B_1^+ $ value. 
 
The reported results may help the community working in MRF in evaluating advantages of deep learning approaches, pushing deep learning approaches and eventually take advantage of them, and pushing fingerprinting pulse sequences design to estimate other MR parameters, such as diffusion, without the limitation of dictionary sizes and of the accuracy of parameters prediction.

\subsection{Background}
Magnetic Resonance Fingerprinting has two main steps to be optimized: an MRI acquisition scheme, which involves pulse sequence design and k-space sampling, and a reconstruction algorithm able to map the acquired signal $S(t)$ to the MR parameter space ($T_1$, $T_2$,$B_0$, $B_1^+$, etc...). 
The latter step is mathematically well described by the concept of the Inverse Transfer Function.

Let us say $\textbf{x}$ is the acquired MRF signal, with $ \textbf{x} \in \mathcal{R}^n$, where $n$ is the size of the acquired signal, and let us say $\textbf{p} \in \mathcal{R}^m$ to be a m-pla in the MR parameter space, where m is the number of parameters such as $T_1$, $T_2$, $B_0$ ect.

Applying a pulse sequence to a system characterized by MR parameters $\textbf{p}$ means that exits a function $g(\textbf{p})$ that maps $\textbf{p}$ into the output signal $\textbf{x}$. Quantitative mapping applies $g^{-1}$ to the acquired signal $\textbf{x}$ to obtain $\textbf{p}$.
Inverse mapping is actually strongly related to pulse sequences design; pulse sequences have usually been designed to obtain the magnetization evolution to be described by an analytic function of the MR parameters, that is to find an analytic form of $g(\textbf{p})$. Examples are the Inversion Recovery sequence for $T_1$ mapping, and Carr-Purcell-Meiboom-Gill sequence for $T_2$ mapping, in which a mono-exponential model or a multi-exponential model are used to describe $g$, and for the multi-exponential case Inverse Laplace Transform is commonly used to obtain $T_1$ and $T_2$ distributions from the acquired signals \cite{G.C.Borgia1998}\cite{UPEN2D}.

In MRF, the pulse sequence is designed to make the magnetization evolution sensitive to multiple parameters at the same time using variable FAs and TRs patterns, making hard to model the transfer function $g(\textbf{p})$ (and $g^{-1}(\textbf{p})$ by consequence) with analytic functions. In such a scenario, template matching with a precomputed dictionary is a way to overcome this lack of knowledge. An outcomes of the transfer function can be simulated for a m-pla of MR parameters $ \textbf{p}_i$ by using Bloch equations, so that an estimated $\hat{g}(\textbf{p}_i)$ can be obtained.  Hence, for a given MRF pulse sequence, a dictionary is built by sampling a set of MR parameters $ \mathcal{P} = \{\textbf{p}_1, \textbf{p}_2,...,\textbf{p}_l\}$,  where $\textbf{p} \in \mathcal{R}^m$.
Then, the simulations of the Transfer Function related to that pulse sequence are stored into a dictionary $\mathcal{D} = \{\hat{g}(\textbf{p}_1), \hat{g}(\textbf{p}_2), ..., \hat{g}(\textbf{p}_l)\}$, where $l$ is the length of the dictionary and  $ \hat{g}(\textbf{p}_i) \in \mathcal{R}^n$.
When a real data is acquired from a voxel, the signal is described by $g(\textbf{p}_{GT})$ where $\textbf{p}_{GT}$ indicates the ground truth MR parameter vector characterizing the voxel tissue. Dictionary based approach maps $g(\textbf{p}_{GT})$ into $\mathcal{D}$ by computing the nearest neighbor between $g(\textbf{p}_{GT})$ and each $\hat{g}(\textbf{p}_i)$ present in $\mathcal{D}$ according to a similarity measure, which usually is the dot product. The main limitation of this approach is related to the a priori design of the dictionary.
The number of entries scales rapidly the more MR parameters are encoded into the dictionary according to the curse of dimensionality \cite{CoD}. Adding a new parameter to be retrieved with MRF leads to an exponential growth of the number of entries to be inserted into the dictionary without affecting resolution. Since both computational and memory usage limitation has to be taken into account, the number of entries should be kept under control.
This could increase sparsity into the dictionary, which can produce high biases in nearest neighbor algorithms \cite{Friedman1997}, especially when they are applied to high dimensional data spaces, such as the case of MRF where usually signals are represented by a 1000 points vector.

With NN approach the limitations of MRF related to dictionaries can be strongly reduced. Indeed, given a training data set $\mathcal{T} = \{\hat{g}(\textbf{p}_1), \hat{g}(\textbf{p}_2), ..., \hat{g}(\textbf{p}_h)\}$, built by Bloch equation simulations, a NN approximation of the inverse transfer function $g^{-1}(S(t))$, defined $\hat{g}^{-1}(S(t))$, can be learned \cite{Hornik1991}.

Once trained, a MRF voxel signal $S(t)$ is the input of the NN model, which applies $\hat{g}^{-1}(S(t))$  to retrieve $\hat{\textbf{p}}$, which is an estimation of $\textbf{p}_{GT}$. Being a regression procedure, the model can correctly predict unknown examples, depending on how well it is trained. This limits the issues of the template matching approach. Moreover, a strong mathematical background exits behind these applications, and in 1989 Hornik demonstrated that standard multilayer feedforward networks are capable of approximating any measurable function to any desired degree of accuracy \cite{Hornik1991}.

\subsection{Background: state of the art of NN applied to MRF}

To our knowledge four works investigated the application of neural networks to MR Fingerprinting. In this subsection the main characteristics of these works have been summarized to highlight the different strategies that have been used so far to apply neural networks to MRF. 

\begin{itemize}
\item Virtue et al. \cite{2017arXiv170700070V} train both a real and complex valued neural networks to estimate $T_1$, $T_2$ and $B_0$ off-resonances from a 500 time points fingerprint given as input, produced by a IR-bSSFP sequence. In particular, in this implementation there are three neural networks, each with the same architecture,  that process in parallel the input fingerprint to predict a MR parameter each.
The neural networks have been trained with simulated data. The main focus of this paper is to investigate if a complex valued neural networks are better than real valued nets. Thus, the NN models have  been implemented to be real valued and complex valued. the The MR parameter space has been sampled using a uniform random sampling. No noise has been added during the training, while white Gaussian noise with SNR = 40 dB has been added for testing the performance. The work has been carried out only with simulated data.

\item Hoppe et al. \cite{Hoppe2017DeepLF} train a convolutional neural network (CNN) with 3 hidden layers for predicting $T_1$ and $T_2$ parameters given a 3000 time points fingerprint as input, produced by a IR-FISP sequence. The network have been trained with simulated data. The MR parameter space has been sampled using a grid sampling. No noise was added during the training neither during the test carried out to asses the performance.

\item Cohen et al. \cite{OuriBoS.2018} train a neural network composed by 2 fully connected hidden layers to predict $T_1$ and $T_2$ parameters. The pulse sequences used in this work are an optimized EPI-MRF pulse sequence \cite{cohen_EPI}, composed by 25 pulses, and IR-FISP pulse sequence with a sliding-window reconstruction \cite{MRF-sliding-window} to reduce undersampling artefacts. The neural network has been trained with simulated data, and the MR parameter space has been sampled in a grid way. White Gaussian noise with zero mean and 1\% standard deviation was added during the training procedure.
The work presents tests both on simulated data and on real data.

\item Balsiger et al. \cite{SpatioTemporal_CNN} train a CNN to predict $T_1$, $T_2$ and $PD$ using an optimized IR-FISP sequence and giving a 4D spatio-temporal patch as input. In particular, the authors want to take advantage of the neighbor pixels during reconstruction to avoid noisy reconstructed parameter maps. With this aim they design the NN application to process $5\times5$ patches in order to predict the MR parameter of the fingerprint in the center of the patch.
The training phase is carried out using real data: six brains where scanned with MRF as well with standard protocols to obtain ground truth maps. A leave-one-out cross-validation has been used to train and test the network performance. 

 The aim of our work is to further investigate the application of NN to perform MRF parameter estimation. For this reasons, both IR-FISP and IR-bSSFP MRF sequences have been tested, as well different training strategies to train the NN models. In particular an accurate study regarding how adding noise during training can affect the generalization performance has been carried out using white Gaussian noise as source of noise.
 Trainig and test phases have been carried out with simulated data.
 
 The NN approach in the present study is used to perform a pixel wise parameter prediction given a MRF signal. Hence, no convolutional layers have been inserted in the NN models. Convolutional neural networks are meant to take advantage of local connections and correlations in the processed signals \cite{LeCunBengioHinton2015}. Moreover, CNN learns filters, meaning that the networks learn to identify a feature independently by its positions within the signal. In case of MRF pixel-wise processing, those characteristics are less meaningful, while are clearly significant for the method used in \cite{SpatioTemporal_CNN}.

\end{itemize}

\section{Methods}

\subsection{MRF pulse sequences and simulations \label{sec:MRF_pulse_sequences}}
\begin{figure}[hbtp]
\subfigure[MRF FISP sequences \label{Fig::FISPs}]{\includegraphics[width=0.45\textwidth]{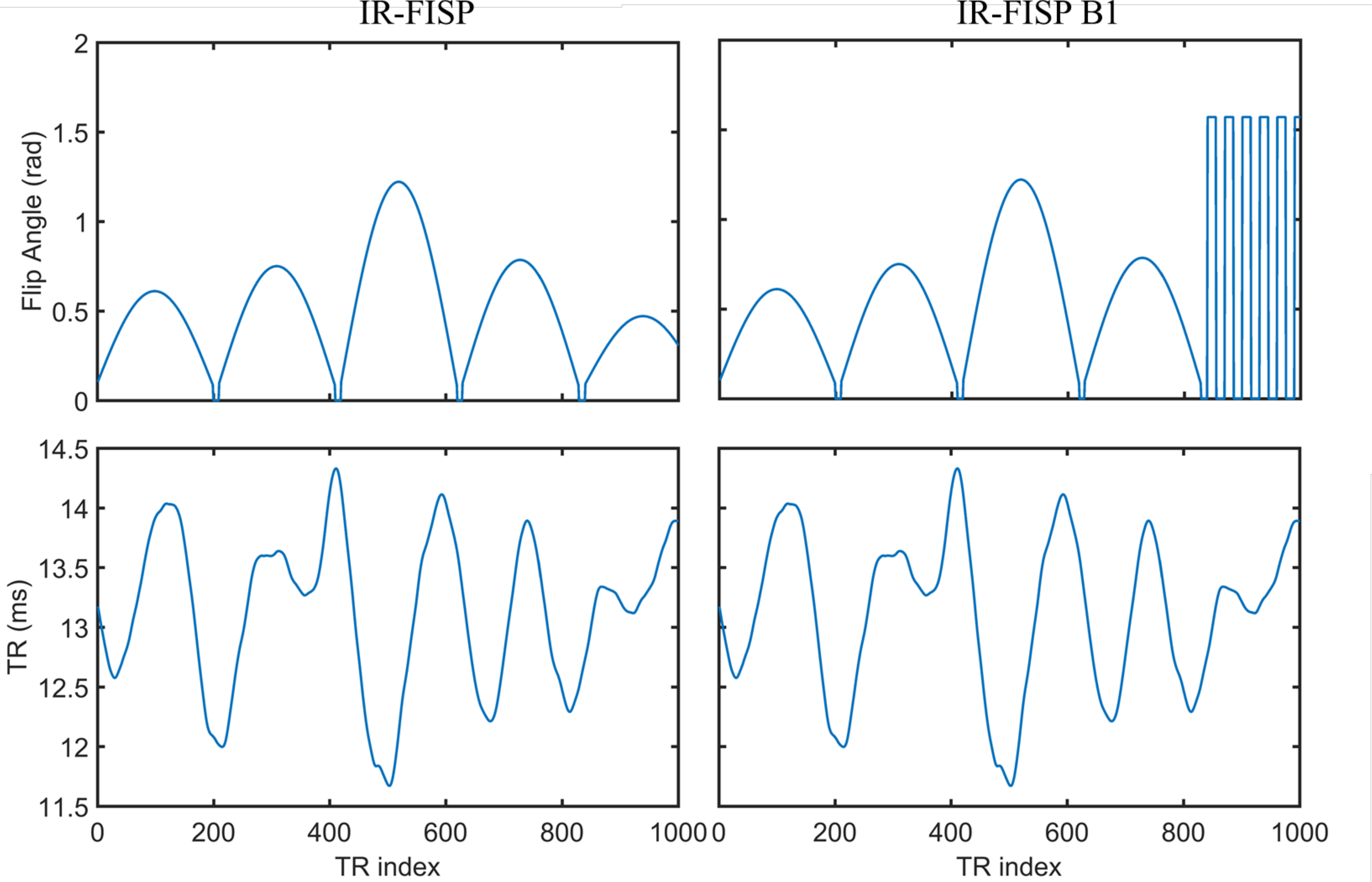}}
\subfigure[MRF bSSFP sequences \label{Fig:bSSFPs}]{\includegraphics[width=0.45\textwidth]{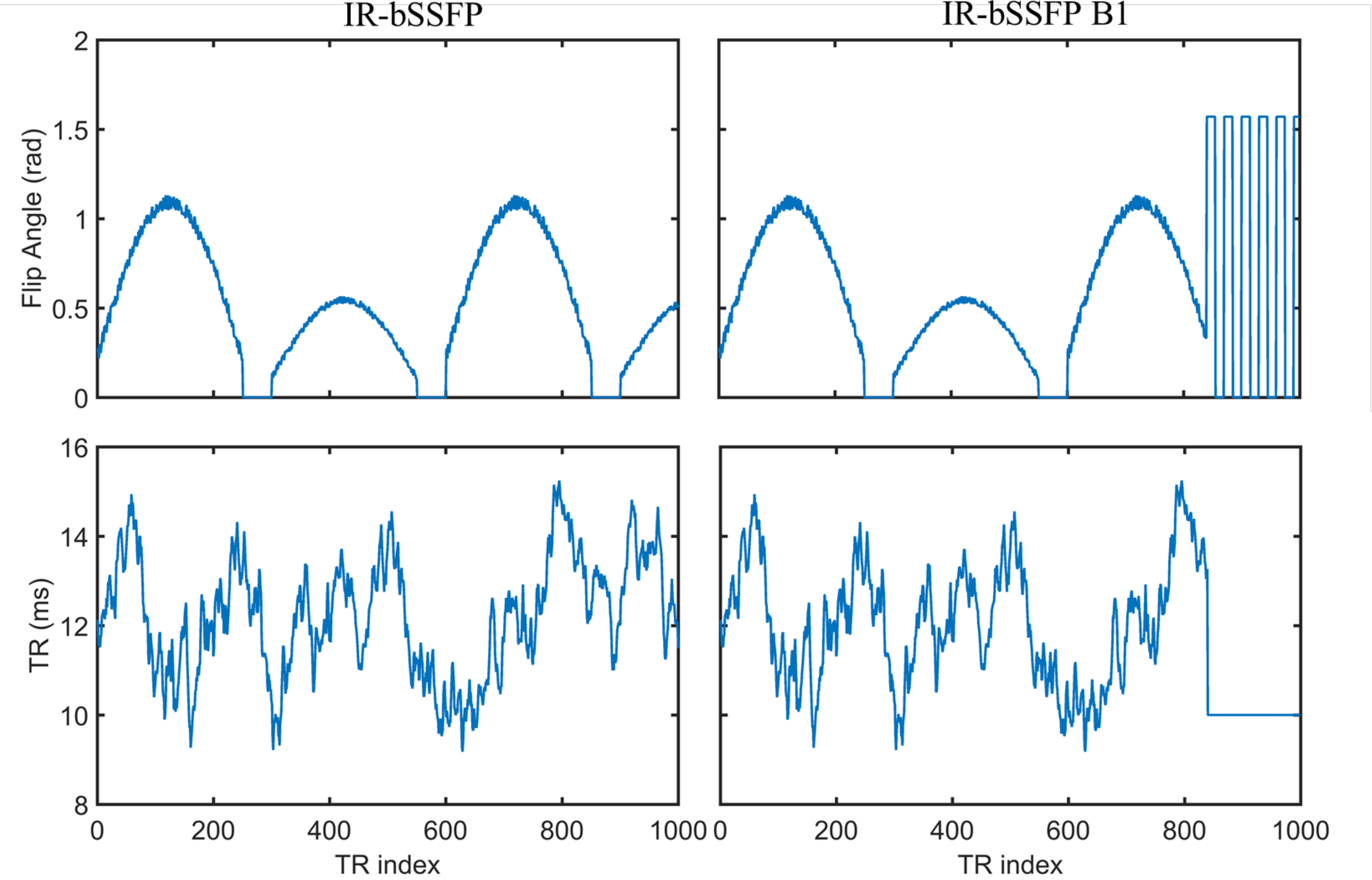}}
\caption{MRF pulse sequences used for generating the synthetic signals. \label{Fig:sequences}}
\end{figure}

The simulations have been performed using four MRF pulse sequences: the MRF IR-FISP pulse sequence as described in \cite{YunDanNicoleEtAl2014}, which encodes $T_1$ and $T_2$, and its variant which accounts for $B_1$ inhomogeneity adding flip angles abrupt changes as described in \cite{BuonincontriSawiak2015}, labeled IR-FISP $B_1$; the MRF IR-bSSFP as described in the original MRF article \cite{Nature2013}, which encodes $T_1$, $T_2$ and $B_0$ off resonance, and a new modification here proposed, labeled IR-bSSFP $B_1$, to take into account for $B_1$ field inhomogeneities. The latter sequence consists in adding abrupt changes at the end of the IR-bSSFP sequence, following the same criterion used for the IR-FISP $B_1$ sequence.
Flip angle and time repetition patterns are depicted in Fig. \ref{Fig:sequences}.

The simulations have been carried out using MATLAB (MathWorks), where Block equations have been used for IR-bSSFP type sequences, whereas the Extended Phase Graph (EPG) algorithm \cite{EPG} has been used for IR-FISP type sequences simulations.

\subsection{Noise}
In this work, noise affecting MRF signals has been considered as complex white Gaussian noise where the variance of the noise is expressed in terms of Signal to Noise Ratio (SNR).
SNR has been defined in two ways as reported in equations \ref{eq:SNR_1} and \ref{eq:SNR_2}, where $P_{signal}$ and $P_{noise}$ represent the power of the MRF signal and the noise, respectively, and $A_{signal}$ represents the signal intensity of the MRF signal and $\sigma_{noise}$  represents the standard deviation of the noise. It is worthy to point out that equation \ref{eq:SNR_2} expresses SNR in dB, but in this work all SNRs are expressed in the linear scale by applying the proper conversion. Which definition of SNR is used during the different experiments of this work is explitly pointed out in the corresponding sections.
\begin{equation}
SNR = \frac{\sigma^2_{signal}}{\sigma^2_{noise}} =  \frac{P_{signal}}{P_{noise}} \label{eq:SNR_1} 
\end{equation}

\begin{equation}
SNR  =  20\times\log_{10} \left(\frac{A_{signal}}{\sigma_{noise}}\right) \label{eq:SNR_2} 
\end{equation}

White Gaussian noise in MRI simulation is used as the most common noise distribution, and, as shown by Griswald et al \cite{McGivneyPierreMaEtAlDec.2014}, severe k-space under-sampling artifacts in MR Fingerprinting can be well described by Gaussian noise with high variance. Being the simulated MRF signal a complex signal, Gaussian noise is added both to the real and imaginary part of it by assigning half of the noise variance per each channel. This indeed simulates a two channels acquisition, and preserves the Rician distribution when the magnitude of the signal is considered \cite{HAKonSamuel}.

\subsection{Feed Forward Neural Network models}
The Deep Neural Network application has been developed using the Python package Keras with TensorFlow \cite{tensorflow2015-whitepaper} backend.

\begin{figure*}[hbtp]
\centering
\includegraphics[width=1\textwidth]{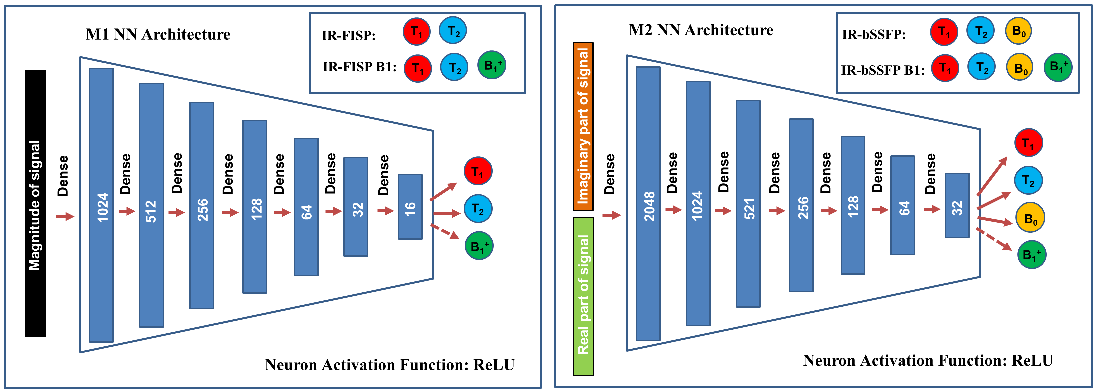}
\caption{NN architectures for model M1 (left box) and model M2 (right box). The two models are fully connected NN with ReLU used as neuron activation function for hidden layers, while linear activation function has been used for the output layer. Blue boxes represents layers and the numbers inside indicate the number of neurons. Model M1 has been used for IR-FISP and IR-FISP $B_1$ pulse sequences, whereas model M2 has been used for IR-bSSFP and IR-bSSFP $B_1$ sequences.\label{Fig:NN_architectures}}
\end{figure*}

Two NN models have been defined, one architecture to process IR-FISP data and another architecture to process IR-bSSFP data. All of them are feed forward nets with 9 fully connected layers, and the Rectified Linear Unit (ReLU) has been used as activation function for the neurons in the firsts 8 layers, while a linear activation function has been chosen for the output layer. 
The NN architectures are reported in Fig.\ref{Fig:NN_architectures} and  described below by indicating the number of neurons per layer:

\begin{itemize}
\item M1: 1000 (input layer) - 1024 - 512 - 256 - 128 - 64 - 32 - 16 - 2  or 3 (output layer);
\item M2: 2000 (input layer)- 2048 - 1024 - 512 - 256 - 128 - 64 - 32 - 3 or 4 (output layer).
\end{itemize}

M1 takes the magnitude of the complex MR signal produced by a IR-FISP or IR-FISP $B_1$ sequence as input, while M2, which is asked to estimate $B_0$, takes in input the concatenated real and imaginary parts of the complex MR signal produced by a IR-bSSFP and IR-bSSFP $B_1$ sequence. The output layer sizes match the number of MR parameters one wants to retrieve from the prediction, and they are:
\begin{itemize}
\item IR-FISP: $T_1$ and $T_2$; 
\item IR-FISP $B_1$: $T_1$, $T_2$ and $B_1^+$;
\item IR-bSSFP: $T_1$, $T_2$ and $B_0$;
\item IR-bSSFP $B_1$: $T_1$, $T_2$, $B_0$ and $B_1^+$.
\end{itemize}

where $B_0$ indicates stationary field off-resonances, and $B_1^+$ indicates $B_1$ excitation field inhomogeneities expressed in terms of correction ratio, being 1 the nominal flip angle excitation. 
 
It is important to point out some of the motivations that have driven the design of the NN model architectures. A bottleneck shape has been selected for two main reasons. The first is avoiding exploding number of model parameters: many layers bring to overfitting, since fully connected layers with large number of neurons increase rapidly the number of model parameters. The second is that a bottleneck shape forces the network to encode more meaningful representations the deeper is the layer considered \cite{LeCunBengioHinton2015}.

Preliminary experiments, not included in this work, where different network sizes, batch sizes and learning rates have been tested have guided to the used architectures. In particular, they showed that the architecture used for IR-bSSFPs sequences works well even for IR-FISPs sequences, whereas the vice-versa does not hold. However, because the smaller is the size of the network, the faster is the computational time, the smallest architecture has been selected for IR-FISPs sequences. 

The training procedure was supervised, using the Mean Squared Error (MSE) between NN estimated parameters and ground truth parameters as loss function, and the Adam algorithm \cite{2014arXiv1412.6980K} has been used for model weights optimization. In particular, 500 epochs with 1000 gradient steps for each epoch have been used, with a fixed batch size of 500. Initial learning rates of $3 \times 10^{-5}$, for model M1, and $1 \times 10^{-4}$, for model M2, have been used. The application has been run on a cluster with 16 dual cores CPUs.

\subsection{Training strategies: training sets, test sets and data augmentation}

\subsubsection{Parameter space sampling, random uniform and grid sampling}
to assess how well the network models learn the ITF depending on the training set distribution, for each pulse sequence two training sets of the same size were generated: a random uniform set, in which the parameter space is sampled using a random uniform distribution; a gridded set, where the parameter space is sampled with a fixed mesh grid.

Table \ref{tab:train_set_summary} summarizes the training set characteristics, where the label \emph{R} refers to the random sampling, whereas label \emph{G} refers to grid sampling.

\begin{table*}[t]
\centering
\begin{adjustbox}{width=1\textwidth}
\small
\begin{tabular}{cccccccc}
\toprule
\textbf{Model} & \textbf{Sequence} & \textbf{Set Label} & \textbf{$T_1$ (ms)} & \textbf{$T_2$ (ms)} & \textbf{$B_0$ (Hz)} & \textbf{$B_1^+$} & \textbf{Count}\\
\midrule
\multirow{2}{*}{M1}         & \multirow{2}{*}{IR-FISP } &  R1  &  [10 $\div$ 4000] & [1 $\div$ 3000] & 0 & 1 & \multirow{2}{*}{75 555}\\
\cmidrule(){4-7}            &                        &  G1  &  [10 $\div$ 4000]$^{*}$ & [1 $\div$ 3000]$^{*}$ & 0 & 1 & \\
\midrule 
\multirow{2}{*}{M1}         & \multirow{2}{*}{IR-FISP $B_1$} &  R2  &  [10 $\div$ 4000] & [1 $\div$ 3000] & 0 & [0.5 $\div$ 1.5] & \multirow{2}{*}{164 475}\\
\cmidrule(){4-7}            &                        &  G2  &  [10 $\div$ 4000]$^{**}$ & [1 $\div$ 3000]$^{**}$ & 0 & [0.5 $\div$ 1.5]$^{**}$ & \\
\midrule
\multirow{2}{*}{M2}         & \multirow{2}{*}{IR-bSSFP} &  R3  &  [10 $\div$ 4000] & [1 $\div$ 3000] & [-400 $\div$ 400] & 1 & \multirow{2}{*}{175 041}\\
\cmidrule(){4-7}            &                        &  G3  &  [10 $\div$ 4000]$^{***}$ & [1 $\div$ 3000]]$^{***}$ & [-400 $\div$ 400]$^{***}$ & 1 &  \\
\midrule  
\multirow{2}{*}{M2}         & \multirow{2}{*}{IR-bSSFP $B_1$} &  R4  &  [10 $\div$ 4000] & [1 $\div$ 3000] & [-400 $\div$ 400] & [0.5 $\div$ 1.5] & \multirow{2}{*}{396 550}\\
\cmidrule(){4-7}            &                        &  G4  &   [10 $\div$ 4000]$^{****}$ & [1 $\div$ 3000]$^{****}$ & [-400 $\div$ 400]$^{****}$ & [0.5 $\div$ 1.5]$^{****}$ & \\
\midrule
 
\end{tabular}
\end{adjustbox}  
\captionof{table}{Summary of the training set used to train the NN models. For G sets the parameter spaces were sampled using the following grids. *) $T_1$ and $T_2$ where incremented with steps of 10 ms; **) $T_1$ = [10:10:800, 850:50:1000, 1100:100:2000, 2500:500:4000] ms, $T_2$ = [1, 10:10:300, 350:50:1000, 600:100:1000, 1500:500:3000] ms and $B_1^+$ = [0.5:0.02:1.5]; ***) $T_1$  = [10, 20:20:500, 600:50:1000, 1100:100:2000, 2500:500:4000] ms, $T_2$ = [1, 10:10:500, 550:50:1000, 600:100:1000, 1500:500:3000] ms and $B_0$  = [ -400:50:150, -100:10:-70, -60:2:60, 70:10:100, 150:50:400] Hz; ****) $T_1$  = [10, 20:20:300, 350:50:500, 600:100:1000, 1250:250:2000, 2500:500:4000], $T_2$  = [1, 10:10:300, 350:50:1000, 600:100:1000, 1500:500:3000] ms. $B_0$  = [-400:50:150, -100:10:-70, -60:2:60, 70:10:100, 150:50:400] Hz and  $B_1^+$ = [0.5:0.1:1.5]. \label{tab:train_set_summary}}
\end{table*}

\subsubsection{Data augmentation and preprocessing}

three data augmentation strategies have been tested by training the network models with different noise adding procedures:
\begin{itemize}
\item \emph{W/O Noise}: using no data augmentation during the training, which means that only noise-free examples are fed to the network models, as in reference \cite{2017arXiv170700070V};
\item \emph{Fixed variance}: feeding the networks with noisy inputs affected by noise with $1\%$ standard deviation, equals to a variance of $10^{-4}$, as in reference \cite{OuriBoS.2018}.
This means that the network during the training sees just one fixed SNR given a set of MR parameters;
\item \emph{Variable variance}: feeding the networks with inputs affected by noise with different variances, which are expressed in terms of SNRs. For each training batch, a vector of SNRs is generated by randomly sampling the SNR values in the range 2 to 100 so that the batch contains data with different SNRs, and the variances of the noises to add to the batch examples have been estimated using equation \ref{eq:SNR_1}.
\end{itemize}

The data augmentation step is done on-line. Hence, given a training set, no new data have to be stored neither in hard neither in flash memories to carry out the three data augmentation strategies. After the data augmentation step, each input is then normalized to have norm equals to 1 to generalize the NN model.

For each model, the performance in predicting a MR parameter has been estimated in terms of Mean Absolute Percentage Error (MAPE) of the predicted parameter, evaluated on a test set composed by 30,000 fingerprints generated using the same pulse sequence as in training procedure, and sampling the parameter space with a random uniform distribution.
To test noise robustness, the described prediction procedure has been repeated giving to each model data with different SNRs = [3, 5, 10, 20, 30, 40, 50, 60, 70 ,80, 90, 100]. Moreover, to assess the variance of the MAPE as a function of the SNR, for each noise level the operation has been repeated 10 times and the standard deviations of the MAPEs has been considered as measure of variance. 
The MAPE is computed following  equation (\ref{eq:mape}), where $\widehat{p_k}$ is the estimated parameter value while $p_k$ is the ground truth parameter value and $N$ is the number of considered examples.

Other error measurements used in this work are the absolute error (AE) and the root mean squared error (RMSE) defined in equations \ref{eq:ae} and \ref{eq:rmse} respectively.

\begin{equation}
MAPE (\%) = \sum_{k=1}^N \vert\frac{\widehat{p_k} - p_k}{p_k}\vert \times 100
\label{eq:mape}
\end{equation}

\begin{equation}
AE = \widehat{p_k} - p_k 
\label{eq:ae}
\end{equation}

\begin{equation}
RMSE = \sqrt{\frac{1}{N}\sum_{k=1}^N \left(\widehat{p_k} - p_k\right)^2} 
\label{eq:rmse}
\end{equation}

\subsection{Numerical Brain Phantom Simulations}
\begin{figure}[hbtp]
\centering
\includegraphics[width=0.5\textwidth]{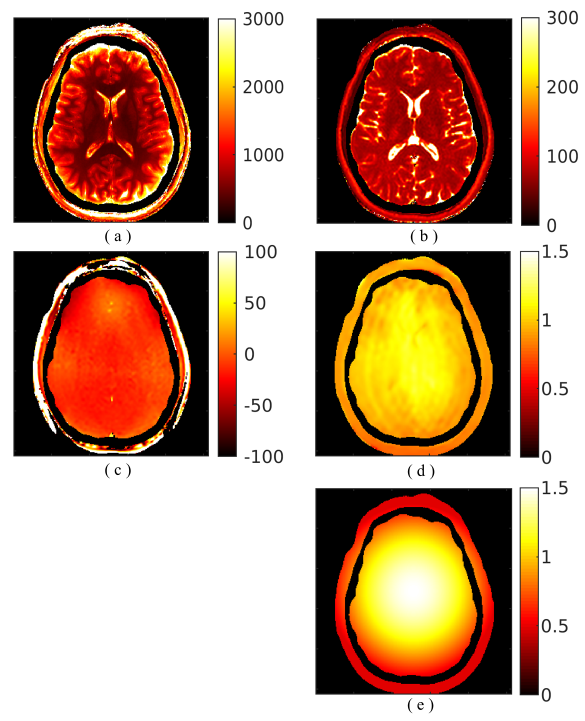}
\caption{Quantitative brain maps computed from the Multi-Modal MRI Reproducibility Resource: (a) $T_1$ (ms), where $T_1$ values from 3000 ms to 4000 ms are showed as equal to 3000 ms); (b) $T_2$ (ms), where $T_2$ values from 300 ms to 2500 ms are showed as equal to 300 ms; (c) $B_0$ (Hz), where  $B_0$ values less than $|500|$ Hz are showed as equal to $|100|$ Hz ; (d) $B_1^+$; (e) synthetic $B_1^+$ (Gaussian shaped).\label{Fig:qMaps_kirby}}
\end{figure}

Realistic $T_1$, $T_2$, $B_0$ and $B_1^+$ maps where obtained by processing real acquisitions downloaded from the Multi-Modal MRI Reproducibility Resource repository. Landman et al \cite{LandmanHuangGiffordEtAl2011} performed MRI acquisitions with standard quantitative protocols at 3 T: variable flip angle (VFA) imaging for $T_1$ mapping; double echo time imaging for $T_2$ mapping; two sequential 2D gradient echo with different echo times for $B_0$ mapping and Actual Flip-Angle Imaging (AFI) for $B_1^+$ mapping. 
The in vivo brain images acquired with these protocols have been then processed with MATLAB to obtain the quantitative maps. In particular, the qMRLab software \cite{JeanFranAoisYeMathieuEtAl} has been used to process the VFA applying $B_1^+$ correction, since $B_1^+$ is known to be a confounding factor to estimate the correct $T_1$ in VFA imaging \cite{MathieuL.NikolaEtAl2017}.
Once obtained, these quantitative maps, reported in Fig. \ref{Fig:qMaps_kirby}, have been used as ground truth to simulate, pixel wise, the MRF acquisition with the four pulse sequences described in section \ref{sec:MRF_pulse_sequences}.
Complex white Gaussian noise was added to the simulated data using the SNR defined in equation \ref{eq:SNR_2} where, in this case, $A_{singal}$ indicates the average signal intensity within a region of the white matter in the first image of the MRF time series.

It is worthy to point out that all the MR parameters have been used to simulate the signals. 
Hence, the $B_1^+$ map has been taken into account both for all the four sequences; in the IR-FISP and IR-FISP $B_1$ the $B_0$ has not be considered because not encoded, while in the IR-bSSFP sequences the $B_0$ is encoded and thus it has been considered in the simulations.
A synthetic $B_1^+$ has also been created with a Gaussian shaped profile exploring a wider range of $B_1$ inhomogeneities as can be found commonly in acquisition with 7 T MRI scanners \cite{BuonincontriSawiak2015}.

For each pulse sequence, the parameter maps have been reconstructed by processing the MRF data with both the trained NN models and using the usual dictionary method based on the dot product. For this latter algorithm, the data sets G1, G2, G3 and G4 have been used as dictionaries for the corresponding pulse sequences.
The MAPE and RMSE have been used as global measure of reconstruction quality, whereas the AE is used as local measure.
To have a complete overview of the performance all three errors have to be assessed.
MAPE gives information about the mean relative error done in parameter estimation, which has the advantage of giving an immediate sense of the global performance, however it could be misleading when very small values are taken into account, such in the case of $B_0$ off-resonances, because small absolute errors give high relative errors.
In such cases, a more reliable global measure is the RMSE, which expresses a global error in the same unit of measure of the considered parameter. 
It is also interesting to check a measure of the local error as the AE, because can be used to assess if the error has a noise-like distribution in the image of it shows strucutres.

\section{Results and Discussion}

\subsection{Training data distribution: Grid Vs Random sampling}
\begin{figure*}[hbtp]
\centering
\includegraphics[width=1\textwidth]{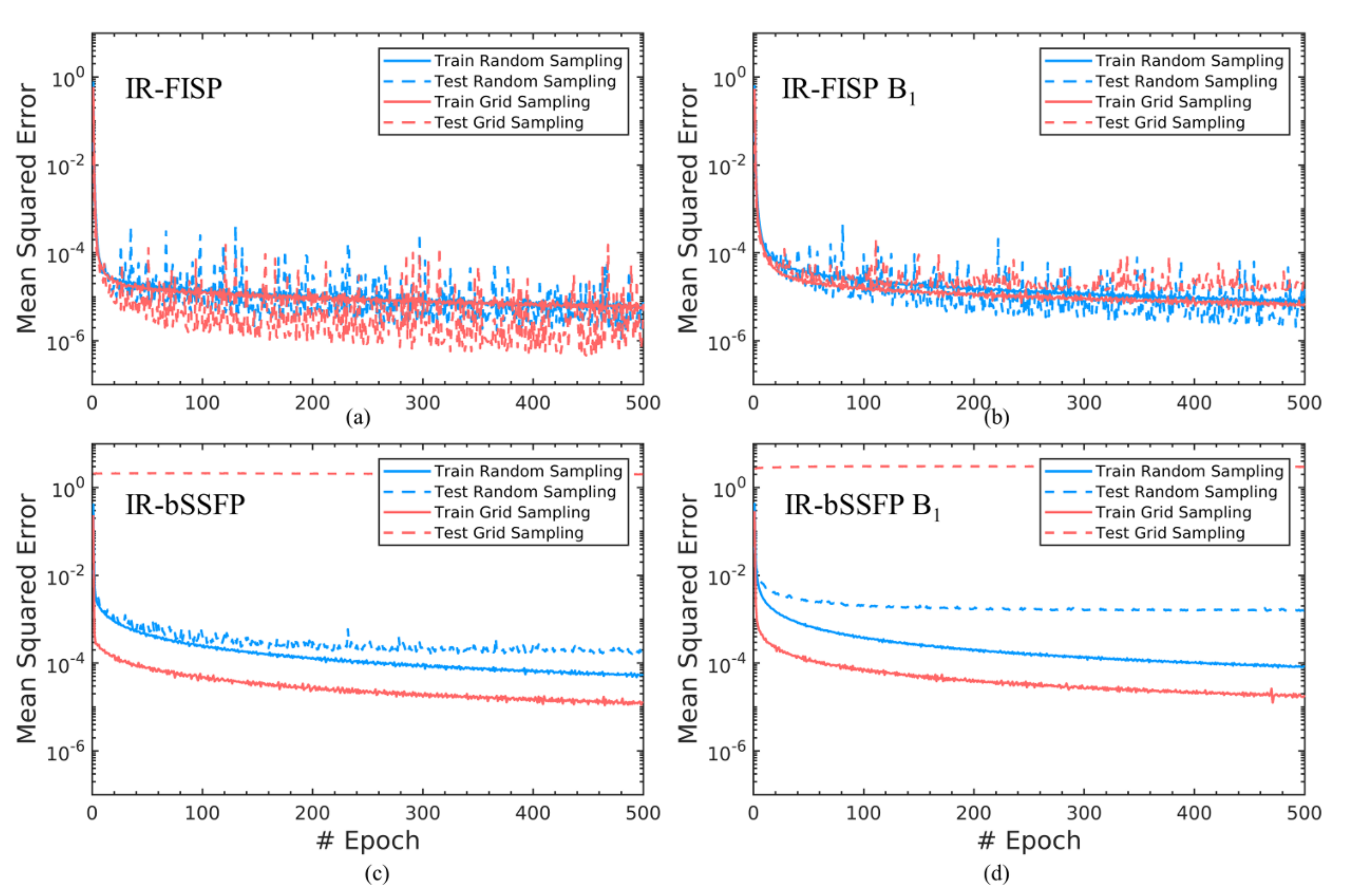}
\caption{Comparison of the training and test loss functions for the four NN models trained with random sampled and grid sampled training data sets.\label{Fig:NoNoie_trLosses}}
\end{figure*}

The results of the training phases of the NN models, fed with noiseless examples with random and grid sampling are summarized in Fig. \ref{Fig:NoNoie_trLosses}, where training and test losses are plotted as function of the training epoch.

\begin{figure*}[hbtp]
\centering
\subfigure[\label{Fig:M3_G}]{\includegraphics[width=1\textwidth]{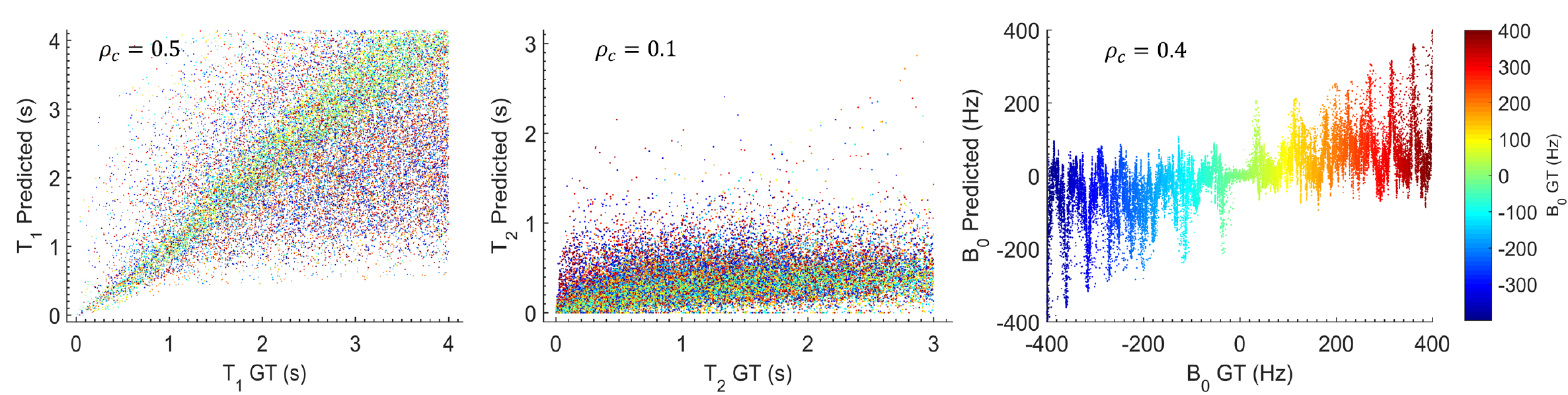}}
\subfigure[\label{Fig:M3_R}]{\includegraphics[width=1\textwidth]{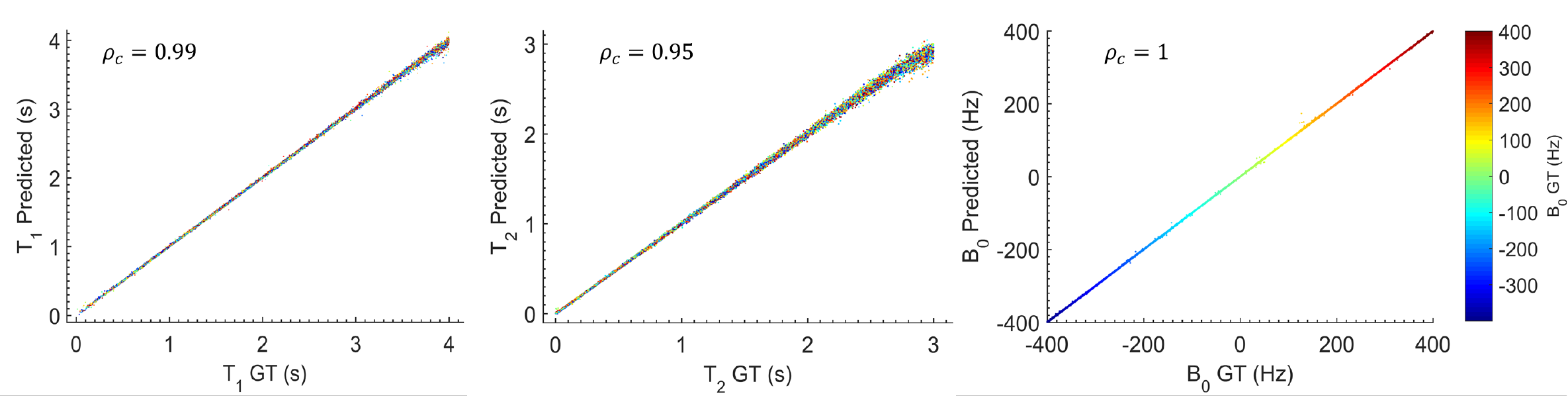}}
\caption{Predicted MR parameters against Ground Truth (GT) parameters using NN model M2 trained with IR-bSSFP MRF examples taken from sets G3 (grid sampling) and R3 (random sampling). Lin's concordance coefficients are also reported as quantitative measures of agreement. Note that the reported predictions refer to test data without added noise. \label{Fig:M3_NoNoise}}
\end{figure*}

For model M1, which processes IR-FISPs type data, there is a strong overlap of the training and test losses as clearly visible looking at Fig. \ref{Fig:NoNoie_trLosses} (a) and (b), indicating an overall well training phase. 
On the contrary, for model M2, which processes IR-bSSFPs type data, the grid training losses are about one order of magnitude lower than the random counterparts, but test losses do not improve during the NN training as clearly visible looking at Fig. \ref{Fig:NoNoie_trLosses} (c) and (d), indicating the NN is overfitting the training data set. 
Hence, model M2, trained with examples sampled in a grid way, overfits the training set.
In figure \ref{Fig:M3_NoNoise} the ground truth MR parameters of the test set against those predicted by the NN model, when trained with randomly and grid sampled IR-bSSFP examples, are shown to appreciate the differences in the performance in estimating the MR parameters. Lin's concordance coefficient \cite{lin_ccc}  between predicted and GT values has been reported as quantitative measure of agreement. The same kind of results holds for IR-bSSFP $B_1$ sequence (not shown in the work).  

Although the grid is a reasonable way to built dictionaries to be used with the classical approach of MRF, indeed it gives information about the accuracy one can expect from the matching algorithm, it is not well performing in train NN in case of IR-bSSFP type sequences ($\rho_c$ is always less than 0.5).
The grid introduces a strong and wrong \emph{a priori} about the real world, which can affect the generalization capacity of the NN, as in the case of IR-bSSFP type pulse sequences. 

\subsection{Noise robustness: learning less to learn better}

\begin{figure*}
\centering
\includegraphics[width=1\textwidth]{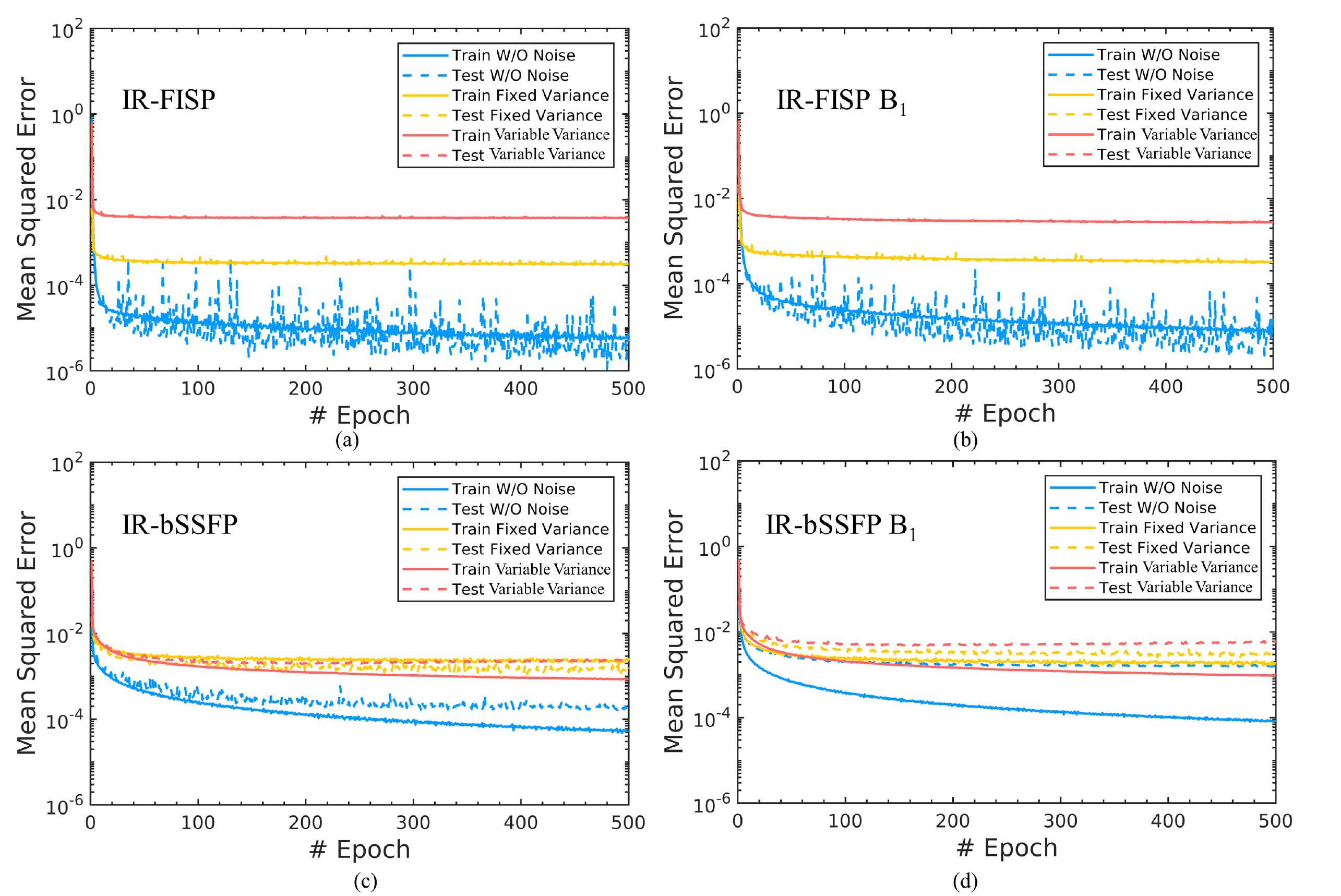}
\caption{Training and test loss functions for the NN models trained with different data augmentation strategies: without noise, \emph{Fixed variance} and \emph{Variable variance}.\label{Fig:Noise_losses_random}}
\end{figure*}

The training and test losses of the NN models trained with different data augmentation strategies are reported in Fig. \ref{Fig:Noise_losses_random}.
It is worthy to notice that for each model, the test loss, where the MSE is used as loss function, is checked at the end of each epoch by asking the model to predict the MR parameters for few never seen examples which however have the same noise adding strategy set for the training step.
The results showed in Fig. \ref{Fig:Noise_losses_random} demonstrate that the best training is reached when the W/O noise data augmentation strategy is used, since both training and test losses reach the lowest values at the end of training, whereas the worst condition for both training and test losses at the end of the 500 training epochs is when the \emph{Variable variance} strategy is used.
Moreover, looking at the gap between training and test losses, model M1 does not have gaps between training and test loss regardless to the data augmentation strategy used, while model M2 always has a gap between the two losses, indicating overfitting.

\begin{figure*}
\centering
\includegraphics[width=1\textwidth]{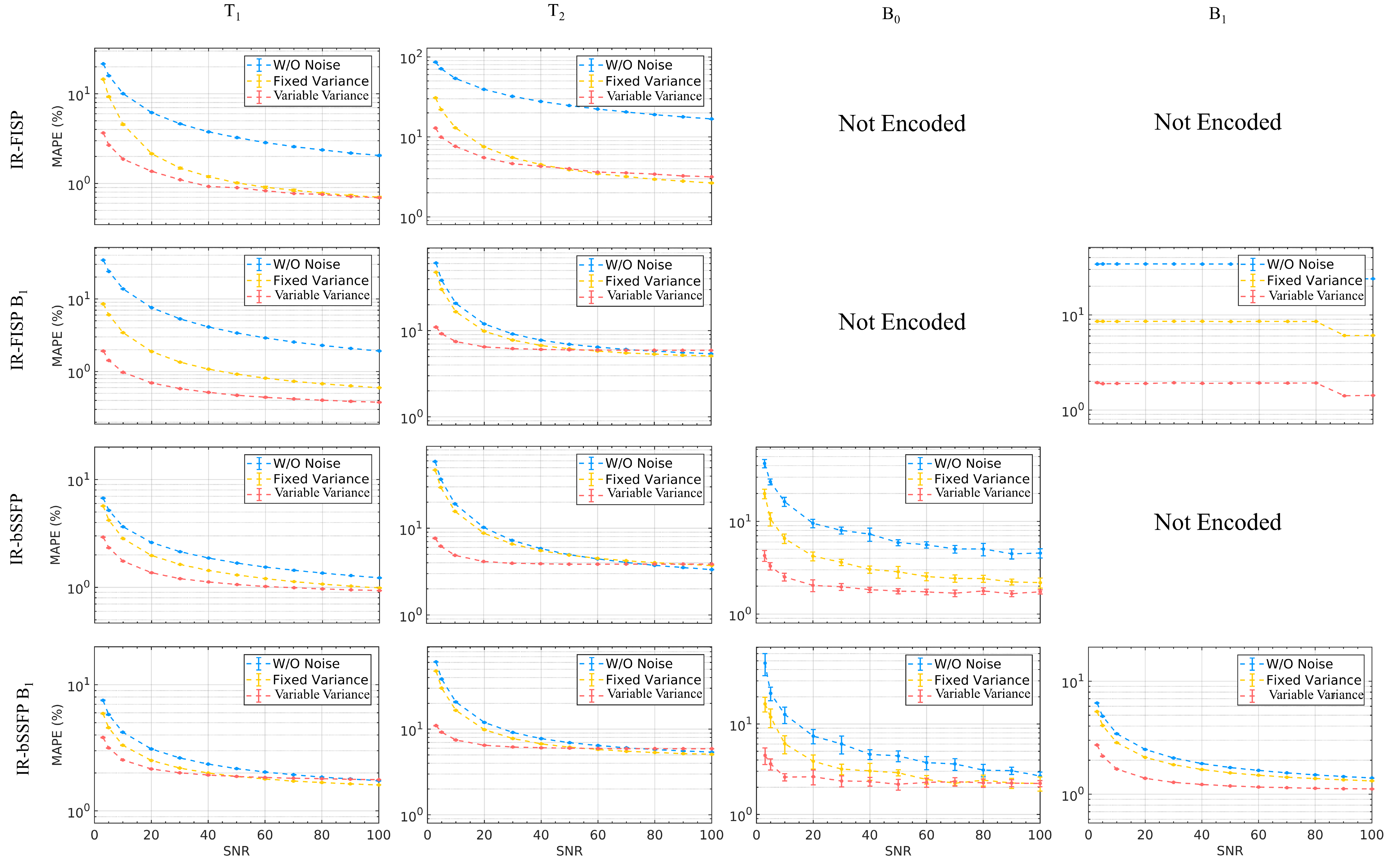}
\caption{Comparison of the parameter mean absolute percentage errors (MAPEs) evaluated on test sets for the NN models trained with different data augmentation strategies (best view on a screen).\label{Fig:Noise_test_mape}}
\end{figure*}

The MAPE of parameters evaluated on test sets for different data augmentation strategies as a function of the SNR are reported in Fig. \ref{Fig:Noise_test_mape}. The error bar indicates two standard deviations of the MAPE, which comes from 10 repetitions of the prediction procedure on noisy data.
Although Fig. \ref{Fig:Noise_losses_random} showed \emph{Variable variance} to be the worst performing strategy during training, it is the most robust to noise among the three.
Looking at Fig. \ref{Fig:Noise_test_mape} one can appreciate the significant improvement in high noise level robustness using this latter strategy for $T_2$ parameter in particular.
For all the models, $T_2$ MAPE never goes over 12\% and decreases rapidly under 10\%, while MAPE for the other parameters is always lower than 5\%. The explanation for this behavior is straightforward: with the \emph{Variable variance} strategy, the NNs get to see more diverse data, and improve generalization.
On the contrary, with the \emph{Fixed variance} strategy the model eventually overfits for data with high SNRs, and, as expected, with the \emph{W/O Noise} strategy the models perform the worst in terms of noise robustness. 
Moreover, since the noise adding happens during the training, and no new data are stored neither in hard and flash memory the \emph{Variable variance} strategy does not affect memory usage efficiency.
Finally, the standard deviation values are in the order of $10^{-2}$, which make them barley visible in the plots, indicating all models show low variance for experiment repetitions.

\begin{figure}
\centering
\includegraphics[width=0.5\textwidth]{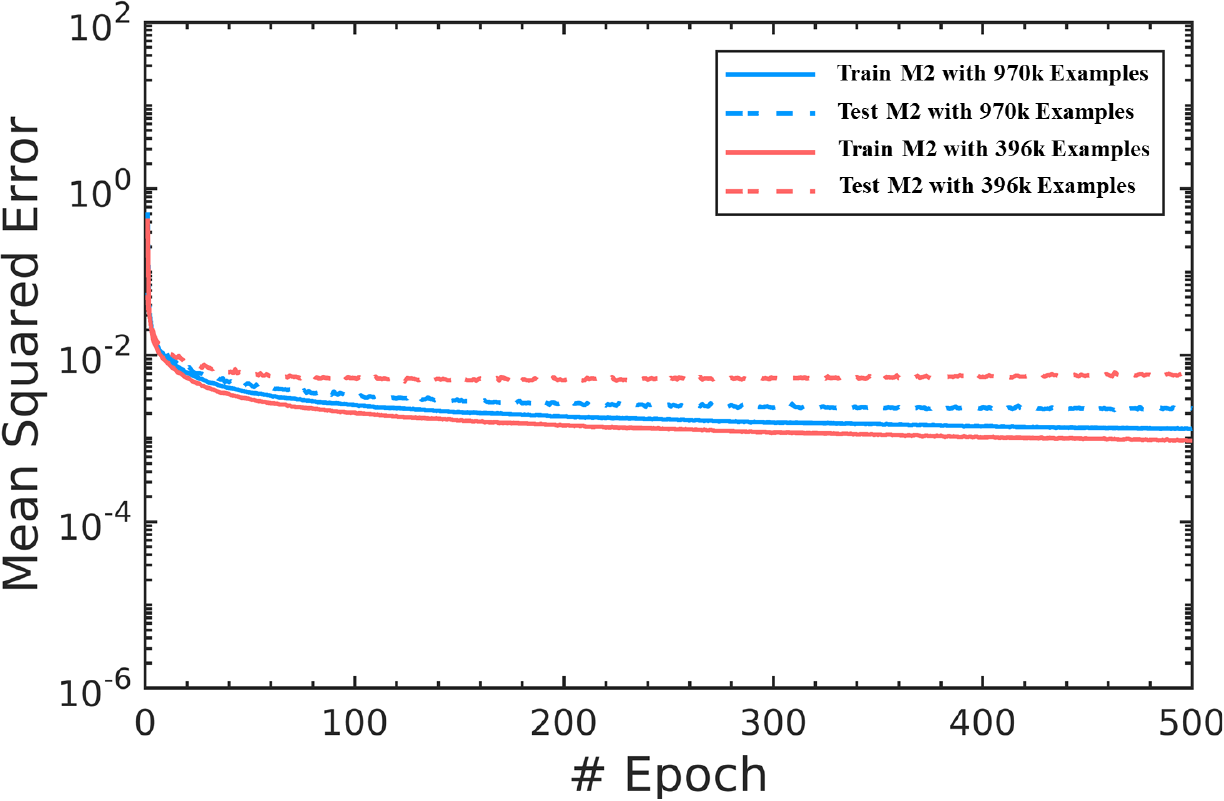}
\caption{Training and test loss functions for the NN model M2 trained with 396 550 training examples and 970 000 training examples.\label{Fig:Loss_M4s}}
\end{figure}

Figure \ref{Fig:Noise_test_mape} shows that for IR-FISP type sequences there is no bias between training and test losses, whereas there is a consistent bias between them for IR-bSSFP type sequences.
In general a difference in training and test losses is expectable, in this case we are testing the models with data coming from the same distribution, although they have never been taken into account for weights optimization during the training phase.
Hence, one could expect training and test losses decrease closely together.
A behavior like the one enhanced here means model M2 is overfitting the training sets and it is likely due to the size of R3 and R4 training sets.
Because the training data are simulated data, a straightforward way to test this hypothesis is to increase the training set size for IR-bSSFP and IR-bSSFP $B_1$ pulse sequences.
A new data set of 1,000,000 examples has been created using the IR-bSSFP $B_1$ sequence. The 97\% of it has been used as training set, while the 3\% as test set for checking test loss during the training phase, which has been performed with the \emph{Variable variance} data augmentation strategy.
In Figure \ref{Fig:Loss_M4s} the comparison between the training and test losses evaluated during the training of NN model M2 with 396 550 and 970 000 examples is reported.
Although the training loss worsens when model M2 is trained with more examples, the gap between training and test loss is considerably reduced and the test loss reaches a lower value than the case in which the model is trained with fewer examples.
Since the test loss is an estimator of the generalization performance of the NN model, it is possible to observe that the generalization of M2 model is increased when more training examples are added.
  
\subsection{Brain maps reconstruction: Neural Networks vs Dictionaries}

\begin{figure*}
\centering
\includegraphics[width=1\textwidth]{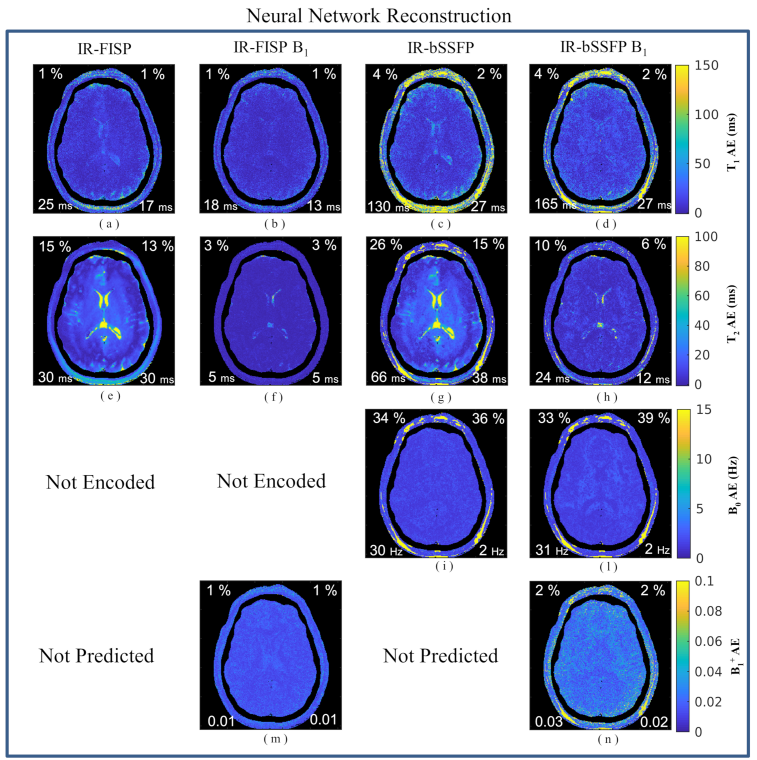}
\caption{Absolute error maps between NN reconstructed and ground truth parameter maps. Global error estimators are reported on the four corners of each image: upper left corner reports the MAPE; upper right corner reports MAPE evaluated without considering the scalp region; lower left corner reports the RMSE while lower right corner reports RMSE evaluated without considering the scalp region.\label{Fig:AE_maps_NN}}
\end{figure*}

\begin{figure*}
\centering
\includegraphics[width=1\textwidth]{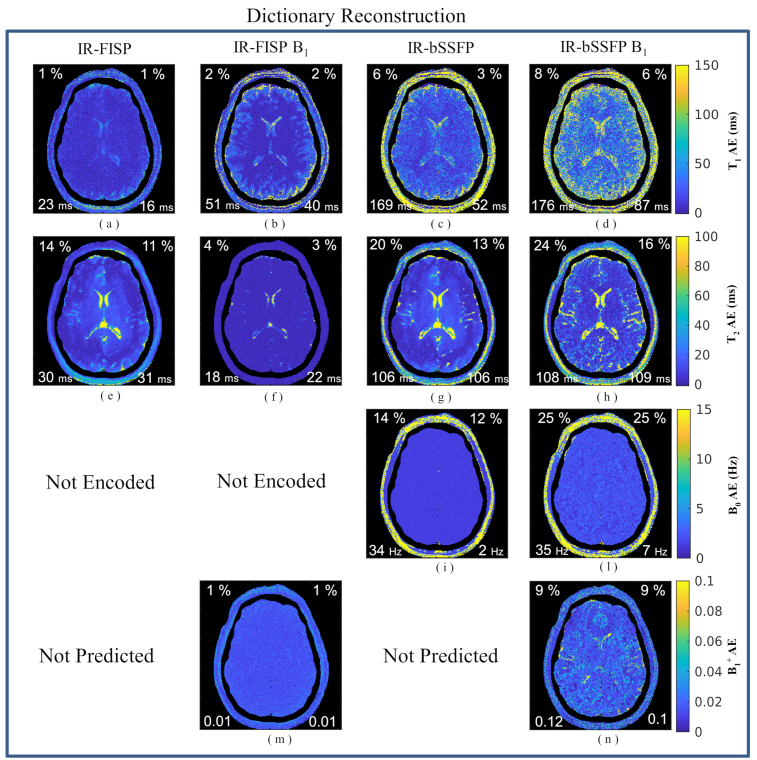}
\caption{Absolute error maps between dictionary reconstructed and ground truth parameter maps. Global error estimators are reported on the four corners of each image: upper left corner reports the MAPE; upper right corner reports MAPE evaluated without considering the scalp region; lower left corner reports the RMSE while lower right corner reports RMSE evaluated without considering the scalp region\label{Fig:AE_maps_D}}
\end{figure*}

In figure \ref{Fig:AE_maps_NN} and Fig. \ref{Fig:AE_maps_D} the absolute error maps for each parameter relative to NN reconstruction and the dictionary matching are reported respectively. They have been evaluated by computing, pixelwise and for each parameter map, the absolute error between the ground-truth parameter value and the predicted value. 
It is worthy to re-mark that the reconstruction has been performed with noisy data with SNR equals to 5.
For each map, four global error estimators have been reported on the corners of the images:
\begin{itemize}
\item upper left corner reports the MAPE between the ground-truth pixel values and those reconstructed;
\item upper right corner reports the MAPE between the ground-truth pixel values and those reconstructed, considering the brain without the scalp;
\item lower left corner reports the RMSE between the ground-truth pixel values and those reconstructed;  
\item lower right corner reports the RMSE between the ground-truth pixel values and those reconstructed, considering the brain without the scalp.
\end{itemize}

Overall the NN approach performs better than the dictionary approach.
This is particularly evident considering the IR-bSSFP $B_1$ sequence, where all the error estimators are lower than those characterizing the dictionary matching.
This difference is even more evident if one looks at the error estimators evaluated without considering the scalp, where the $B_0$ off resonances go up to +/- 500 Hz, which are out of the range used for training the NN model M2, and, as one would expect, this affects parameters estimation accuracy. 
The scalp region is of poor clinical interest as most of the information is in the brain region.
For this reason, the global error estimators have been computed also without considering the scalp. In the latter case the error estimators show a better agreement with ground truth values even if the SNR is very low. 
It is worthy to point out that MAPE is a poor measure of quality for $B_0$ estimation, because small absolute errors made by the NN models are high percentage errors, as highlighted by looking at the RMSEs, which are in the order of 2 Hz.

Considering the IR-FISP the two approaches have similar performance probably because the number of the examples used for build the dictionaries was high enough to ensure a proper MR space sampling also for the dictionary approach.
For the IR-FISP $B_1$, instead, the NN approach performs better in estimating $T_1$ and $T_2$. MAPEs and RMSEs evaluated using the NN recontruction are less or equal than MAPE and RMSE evaulated using dictionary reconstruction.
Analyzing more in detail Fig. \ref{Fig:AE_maps_NN}, one can appreciate how not taking into account for $B_1^+$ field inhomogeneities drives to drop in accuracy in parameter estimation (Fig. \ref{Fig:AE_maps_NN}e, Fig. \ref{Fig:AE_maps_NN}g, Fig. \ref{Fig:AE_maps_D}w and \ref{Fig:AE_maps_D}g), being $T_2$ the parameter mostly affected. In particular, $T_2$ MAPE increases from 2.5\% to 13\% (Fig. \ref{Fig:AE_maps_NN}f and Fig. \ref{Fig:AE_maps_NN}e respectively), when $B_1^+$ is not taken into account during network training for IR-FISP sequences.
The same drop in $T_2$ accuracy happens for IR-bSSFP sequences (Fig. \ref{Fig:AE_maps_NN}g and Fig. \ref{Fig:AE_maps_NN}h).
Considering Fig. \ref{Fig:AE_maps_D}, one can conclude that taking into account for $B_1^+$ field inhomogeneities does not improve parameter estimations for IR-bSSFP sequences. On the contrary, errors are higher when $B_1^+$ field inhomogeneities are predicted in the dictionary, a trend that is opposite to what obtained using NN approach. This behaviour is because the number of entries of the dictionary is not enough to guarantee an accurate sampling of the 4D MR parameter space. It is an example of the  curse of dimensionality of nearest neighbor algorithms.

\begin{figure*}
\centering
\includegraphics[width=1\textwidth]{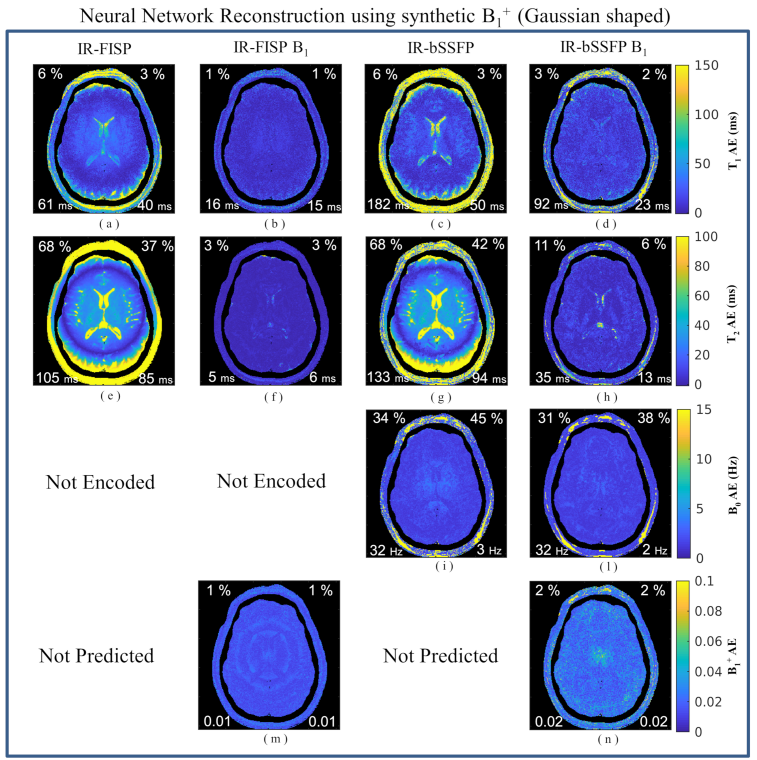}
\caption{Absolute error maps between NN reconstructed and ground truth parameter maps with Gaussian shaped $B_1^+$ field inhomogeneities. Global error estimators are reported on the four corners of each image: upper left corner reports the MAPE; upper right corner reports MAPE evaluated without considering the scalp region; lower left corner reports the RMSE while lower right corner reports RMSE evaluated without considering the scalp region. \label{Fig:AE_maps_gauss_NN}}
\end{figure*}

\begin{figure*}
\centering
\includegraphics[width=1\textwidth]{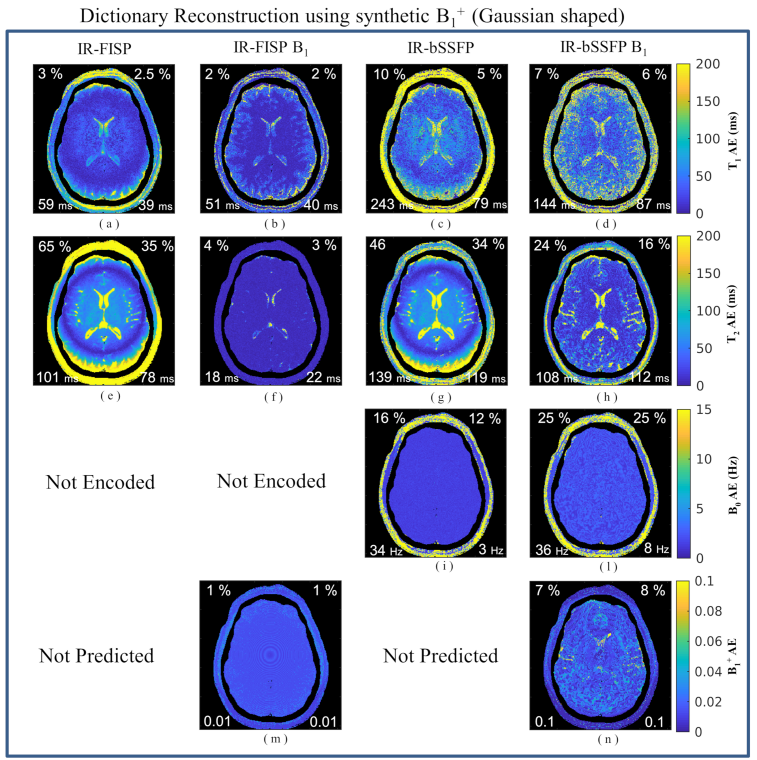}
\caption{Absolute error maps between dictionary reconstructed and ground truth parameter maps with Gaussian shaped $B_1^+$ field inhomogeneities. Global error estimators are reported on the four corners of each image: upper left corner reports the MAPE; upper right corner reports MAPE evaluated without considering the scalp region; lower left corner reports the RMSE while lower right corner reports RMSE evaluated without considering the scalp region.\label{Fig:AE_maps_gauss_D}}
\end{figure*}

All these considerations are reinforced by looking at Fig. \ref{Fig:AE_maps_gauss_NN} and Fig. \ref{Fig:AE_maps_gauss_D}, where the same measurements have been repeated while considering not the original $B_1^+$ map, but the synthetic Gaussian shaped $B_1^+$ map.
IR-bSSFP $B_1$ sequence is a clear example that highlights the difference between NN and dictionary approaches when more than two or three parameters are encoded in the simulations.
Indeed, even if the same number of examples have been given to the two approaches, the dictionary method suffers from the coarse resolution of the MR parameter space sampling, whereas the NN, trained with the same number of examples, is able to predict parameter values with relative high accuracy.

The NN reconstruction lasts about 1.5 s for IR-FISP type sequences and 4 s for IR-bSSFP type sequences to elaborate 35,000 pixels.
No significant time increase has been observed when estimating 2 or 3 parameters with the IR-FISP type sequences and 3 or 4 parameters with the IR-bSSFP type sequences.
This is because the only difference in NN models for IR-FISP type sequences, and NN for IR-bSSFP type sequences, is in the output layers, and this just adds a small number of operations. 
On the contrary, with dictionary matching the computational time strongly scales with the number of parameters, because the size of the dictionary increases, being 10 s, 17 s, 77 s and 160 s for IR-FISP, IR-FISP $B_1$, IR-bSSFP and IR-bSSFP $B_1$ respectively. Overall the NN reconstruction is more efficient both in memory usage, since it does not need to store any dictionary in the RAM memory, both in computational time.

\section{Conclusions}
In this work a deep neural network applied to MRF data has been proposed, tested both for prediction accuracy and noise robustness by means of simulated data. 
Intrinsic limitations of dictionary-based approaches for MRF, and how NNs approaches can overcome them, have been discussed in details.
In particular, it has been studied how to better design parameter space sampling and how to achieve better noise robustness during NN training. 
To demonstrate the generalization of the proposed NN, four different MRF pulse sequences have been considered: the IR-FISP sequence and its variant to account for $B_1^+$ field inhomogeneities, the original IR-bSSFP and a newly proposed variant to account for $B_1^+$ field inhomogeneities. 
The results have demonstrated the low accuracy of grid sampling to build training data sets, since this method introduces a strong bias due to its regularity, and this leads the NN to overfit, as shown for IR-bSSFP type sequences. 
In fact, random uniform sampling  has been found to perform better in spanning the parameter space for the IR-bSSFP type sequence, whereas not significant differences have been found for the IR-FISP type sequence between grid and random sampling.

Then, three different data augmentation strategies have been tested in order to promote robustness to white Gaussian noise.
Feeding the network models with noisy data with different noise variances have yielded the best results in terms of noise robustness providing a MAPE lower than 15\% (evaluated on a wide range of parameter values)  even in case of severe noisy data (SNR $=$ 3).
These results exceed those obtained with the data augmentation strategies used in \cite{2017arXiv170700070V} and \cite{OuriBoS.2018}, which have also been tested.

Finally, a comparison between the NN  and the dictionary approaches has been performed using a numerical brain phantom built from real standard quantitative MRI protocols.
The results of this comparison have shown that NN performance is greater or equal to those of dictionary-based approach, depending on the pulse sequence.

A neural network model learns an approximation of the inverse transfer function from a set of training examples.
Once trained, the processing is computationally fast and efficient, not requiring a large memory usage for dictionary storage.
To increase parameter estimation accuracy is important to take into account as many scanner artifacts as possible, such as $B_0$ and $B_1^+$ inhomogeneities.
With the classical dictionary-based approach for MRF, each added parameter results in an exponential increase of dictionary entries, making it difficult to add other relevant MR parameters, such as diffusion. 
These results show the advantages of NN approaches in MR Fingerprinting parameters estimation. 
Thousands of training examples might be required to learn an accurate and robust ITF with NN, as shown for IR-bSSFP and IR-bSSFP $B_1$ sequences.
However, because the NN model is trained with batches, the data can be stored in the hard memory, and just few of them loaded into the RAM in each step of the training.
Moreover, because the training examples are synthetic data, one can even make the simulation step within the training pipeline, without the need to store any data even in the hard memory.
The latter step requires the simulation code to be fast and efficient, otherwise it becomes a bottleneck in terms of time needed to train the network model. However, parallel coding is well suited for this kind of problem and can be used to make computation faster.

The above results describe NN approaches as promising for MRF application. Hence, this work may help the community working in MRF to build more trust in deep learning approaches and eventually take advantage of them, and pushing fingerprinting pulse sequences design to add more meaningful MR parameters, such as diffusion, without the limitations due to dictionary size.

\bibliography{biblio.bib}

\begin{thebibliography}{10}

\bibitem{Nature2013}
D.~Ma, V.~Gulani, N.~Seiberlich, K.~Liu, J.~L. Sunshine, J.~L. Duerk, and M.~A.
  Griswold, ``Magnetic resonance fingerprinting,'' {\em Nature}, vol.~495,
  p.~187–192, Mar. 2013.

\bibitem{LemassonPannetierCoqueryEtAl2016}
B.~Lemasson, N.~Pannetier, N.~Coquery, L.~S.~B. Boisserand, N.~Collomb,
  N.~Schuff, M.~Moseley, G.~Zaharchuk, E.~L. Barbier, and E.~L. Christen, ``Mr
  vascular fingerprinting in stroke and brain tumors models,'' {\em Sci. Rep.},
  vol.~6, p.~37071, Nov. 2016.

\bibitem{ChenJiangPahwaEtAl2016}
Y.~Chen, Y.~Jiang, S.~Pahwa, D.~Ma, L.~Lu, M.~D. Twieg, K.~L. Wright,
  N.~Seiberlich, M.~A. Griswold, and V.~Gulani, ``Mr fingerprinting for rapid
  quantitative abdominal imaging,'' {\em Radiology}, vol.~279, no.~1,
  pp.~278--286, 2016.
\newblock PMID: 26794935.

\bibitem{doi:10.1148/radiol.2018180836}
Y.~Chen, A.~Panda, S.~Pahwa, J.~I. Hamilton, S.~Dastmalchian, D.~F. McGivney,
  D.~Ma, J.~Batesole, N.~Seiberlich, M.~A. Griswold, D.~Plecha, and V.~Gulani,
  ``Three-dimensional mr fingerprinting for quantitative breast imaging,'' {\em
  Radiology}, 2018.
\newblock PMID: 30375925.

\bibitem{JinghuaWeihuaMaolinEtAl}
J.~Wang, W.~Mao, M.~Qiu, M.~B. Smith, and R.~T. Constable, ``Factors
  influencing flip angle mapping in mri: Rf pulse shape, slice-select
  gradients, off-resonance excitation, and b0 inhomogeneities,'' {\em Magn.
  Reson. Med.}, vol.~56, no.~2, pp.~463--468, 2006.

\bibitem{BuonincontriSawiak2015}
G.~Buonincontri and S.~J. Sawiak, ``Mr fingerprinting with simultaneous b1
  estimation,'' {\em Magn. Reson. Med.}, vol.~76, pp.~1127--1135, Sept. 2015.

\bibitem{MRF_LW_SM_2018}
D.~Ma, S.~Coppo, Y.~Chen, D.~F. McGivney, Y.~Jiang, S.~Pahwa, V.~Gulani, and
  M.~A. Griswold, ``Slice profile and b1 corrections in 2d magnetic resonance
  fingerprinting,'' {\em Magnetic Resonance in Medicine}, vol.~78, no.~5,
  pp.~1781--1789, 2017.

\bibitem{McGivneyPierreMaEtAlDec.2014}
D.~F. McGivney, E.~Pierre, D.~Ma, Y.~Jiang, H.~Saybasili, V.~Gulani, and M.~A.
  Griswold, ``Svd compression for magnetic resonance fingerprinting in the time
  domain,'' {\em IEEE Trans. Med. Imaging}, vol.~33, no.~12, pp.~2311--2322,
  Dec. 2014.

\bibitem{FLOR_MRF}
G.~Mazor, L.~Weizman, A.~Tal, and Y.~C. Eldar, ``{Low Rank Magnetic Resonance
  Fingerprinting},'' {\em ArXiv e-prints}, Jan. 2017.

\bibitem{MingruiDanYunEtAl2017}
M.~Yang, D.~Ma, Y.~Jiang, J.~Hamilton, N.~Seiberlich, M.~A.Griswold, and
  D.~McGivney, ``Low rank approximation methods for mr fingerprinting with
  large scale dictionaries,'' {\em Magn. Reson. Med}, vol.~79, pp.~2392--2400,
  Aug. 2017.

\bibitem{Hornik1991}
K.~Hornik, ``Approximation capabilities of multilayer feedforward networks,''
  {\em Neural Networks}, vol.~4, no.~2, pp.~251 -- 257, 1991.

\bibitem{2017arXiv170700070V}
P.~Virtue, S.~X. Yu, and M.~Lustig, ``{Better than Real: Complex-valued Neural
  Nets for MRI Fingerprinting},'' {\em ArXiv e-prints}, June 2017.

\bibitem{Hoppe2017DeepLF}
E.~Hoppe, G.~K{\"o}rzd{\"o}rfer, T.~W{\"u}rfl, J.~Wetzl, F.~Lugauer,
  J.~Pfeuffer, and A.~K. Maier, ``Deep learning for magnetic resonance
  fingerprinting: A new approach for predicting quantitative parameter values
  from time series,'' {\em Studies in health technology and informatics},
  vol.~243, pp.~202--206, 2017.

\bibitem{OuriBoS.2018}
O.~Cohen, B.~Zhu, and M.~S. Rosen, ``Mr fingerprinting deep reconstruction
  network (drone),'' {\em Magn. Reson. Med}, vol.~80, pp.~885--894, Apr. 2018.

\bibitem{SpatioTemporal_CNN}
F.~{Balsiger}, A.~{Shridhar Konar}, S.~{Chikop}, V.~{Chandran},
  O.~{Scheidegger}, S.~{Geethanath}, and M.~{Reyes}, ``{Magnetic Resonance
  Fingerprinting Reconstruction via Spatiotemporal Convolutional Neural
  Networks},'' {\em ArXiv e-prints}, July 2018.

\bibitem{YunDanNicoleEtAl2014}
Y.~Jiang, D.~Ma, N.~Seiberlich, V.~Gulani, and M.~A. Griswold, ``Mr
  fingerprinting using fast imaging with steady state precession (fisp) with
  spiral readout,'' {\em Magn. Reson. Med.}, vol.~74, pp.~1621--1631, Dec.
  2014.

\bibitem{G.C.Borgia1998}
G.~C. Borgia, R.~J.~S. Brown, and P.~Fantazzini, ``Uniform-penalty inversion of
  multiexponential decay data,'' {\em J Magn Reson}, no.~132, p.~65â€“77,
  1998.

\bibitem{UPEN2D}
V.~Bortolotti, L.~Brizi, P.~Fantazzini, G.~Landi, and F.~Zama, ``Filtering
  techniques for efficient inversion of two-dimensional nuclear magnetic
  resonance data,'' {\em Journal of Physics: Conference Series}, vol.~904,
  no.~1, p.~012005, 2017.

\bibitem{CoD}
R.~Bellman, {\em Adaptive control processes: a guided tour}.
\newblock Princeton University Press, 1961.

\bibitem{Friedman1997}
J.~H. Friedman, ``On bias, variance, 0/1--loss, and the
  curse-of-dimensionality,'' {\em Data Mining and Knowledge Discovery}, vol.~1,
  no.~1, pp.~55--77, 1997.

\bibitem{cohen_EPI}
O.~Cohen and M.~S. Rosen, ``Algorithm comparison for schedule optimization in
  mr fingerprinting,'' {\em Magnetic Resonance Imaging}, vol.~41, pp.~15 -- 21,
  2017.

\bibitem{MRF-sliding-window}
X.~Cao, C.~Liao, Z.~Wang, Y.~Chen, H.~Ye, H.~He, and J.~Zhong, ``Robust
  sliding-window reconstruction for accelerating the acquisition of mr
  fingerprinting,'' {\em Magnetic Resonance in Medicine}, vol.~78, no.~4,
  pp.~1579--1588, 2017.

\bibitem{LeCunBengioHinton2015}
Y.~LeCun, Y.~Bengio, and G.~Hinton, ``Deep learning,'' {\em Nature}, vol.~521,
  p.~436–444, May 2015.

\bibitem{EPG}
H.~Jürgen, ``Echoes—how to generate, recognize, use or avoid them in
  mr-imaging sequences. part i: Fundamental and not so fundamental properties
  of spin echoes,'' {\em Concepts in Magnetic Resonance}, vol.~3, no.~3,
  pp.~125--143, 1991.

\bibitem{HAKonSamuel}
H.~Gudbjartsson and S.~Patz, ``The rician distribution of noisy mri data,''
  {\em Magn. Reson. Med.}, vol.~34, no.~6, pp.~910--914, 1995.

\bibitem{tensorflow2015-whitepaper}
M.~Abadi, A.~Agarwal, P.~Barham, E.~Brevdo, Z.~Chen, C.~Citro, G.~S. Corrado,
  A.~Davis, J.~Dean, M.~Devin, S.~Ghemawat, I.~Goodfellow, A.~Harp, G.~Irving,
  M.~Isard, Y.~Jia, R.~Jozefowicz, L.~Kaiser, M.~Kudlur, J.~Levenberg,
  D.~Man\'{e}, R.~Monga, S.~Moore, D.~Murray, C.~Olah, M.~Schuster, J.~Shlens,
  B.~Steiner, I.~Sutskever, K.~Talwar, P.~Tucker, V.~Vanhoucke, V.~Vasudevan,
  F.~Vi\'{e}gas, O.~Vinyals, P.~Warden, M.~Wattenberg, M.~Wicke, Y.~Yu, and
  X.~Zheng, ``{TensorFlow}: Large-scale machine learning on heterogeneous
  systems,'' 2015.
\newblock Software available from tensorflow.org.

\bibitem{2014arXiv1412.6980K}
D.~P. Kingma and J.~Ba, ``{Adam: A Method for Stochastic Optimization},'' {\em
  ArXiv e-prints}, Dec. 2014.

\bibitem{LandmanHuangGiffordEtAl2011}
B.~A. Landman, A.~J. Huang, A.~Gifford, D.~S. Vikram, I.~A.~L. Lim, J.~A.
  Farrell, J.~A. Bogovic, J.~Hua, M.~Chen, S.~Jarso, S.~A. Smith, S.~Joel,
  S.~Mori, J.~J. Pekar, P.~B. Barker, J.~L. Prince, and P.~C. van Zijl,
  ``Multi-parametric neuroimaging reproducibility: A 3-t resource study,'' {\em
  NeuroImage}, vol.~54, no.~4, pp.~2854 -- 2866, 2011.

\bibitem{JeanFranAoisYeMathieuEtAl}
J.-F. Cabana, Y.~Gu, M.~Boudreau, I.~R. Levesque, Y.~Atchia, J.~G. Sled,
  S.~Narayanan, D.~L. Arnold, G.~B. Pike, J.~Cohen-Adad, T.~Duval, M.-T. Vuong,
  and N.~Stikov, ``Quantitative magnetization transfer imaging made easy with
  qmtlab: Software for data simulation, analysis, and visualization,'' {\em
  Concepts in Magnetic Resonance Part A}, vol.~44A, no.~5, pp.~263--277, 2015.

\bibitem{MathieuL.NikolaEtAl2017}
M.~Boudreau, C.~L. Tardif, N.~Stikov, J.~G. Sled, W.~Lee, and G.~B. Pike, ``B1
  mapping for bias-correction in quantitative t1 imaging of the brain at 3t
  using standard pulse sequences,'' {\em J. Magn. Reson. Imaging}, vol.~46,
  no.~6, pp.~1673--1682, 2017.

\bibitem{lin_ccc}
L.~I.-K. Lin, ``A concordance correlation coefficient to evaluate
  reproducibility,'' {\em Biometrics}, vol.~45, no.~1, pp.~255--268, 1989.

\end{thebibliography}

\end{document}